\documentstyle[11pt,moriond,epsfig]{article}

\def\Journal#1#2#3#4{{#1} {\bf #2}, #3 (#4)}

\def\NIMA{{ Nucl. Instrum. Methods} A}

\def\PLB{{\em Phys. Lett.}  B}

\def\ZPC{{\em Z. Phys.} C}

\def\kl{K_L}
\def\ks{K_S}
\def\pppo{\pi^0\pi^0\pi^0}
\def\ppo{\pi^0\pi^0}
\def\pppco{\pi^+\pi^-\pi^0}
\def\ppc{\pi^+\pi^-}
\def\pen{\pi^{\pm} e^{\mp} \nu }
\def\pmn{\pi^{\pm} \mu^{\mp} \nu }

\def\be{\begin{equation}}
\def\ee{\end{equation}}
\def\bea{\begin{eqnarray}}
\def\eea{\end{eqnarray}}

\begin{document}
\vspace*{4cm}
\title{$|V_{us}|$ AND $K_S$ DECAYS FROM KLOE}

\author{\footnotesize THE KLOE COLLABORATION\footnote{
The KLOE Collaboration:
A.~Aloisio,
F.~Ambrosino,
A.~Antonelli, 
M.~Antonelli,
C.~Bacci,
G.~Bencivenni,
S.~Bertolucci,
C.~Bini,
C.~Bloise,
V.~Bocci
% ...
}}

\vskip 0.5cm
\author{presented by G. LANFRANCHI}

\address{INFN Laboratori Nazionali di Frascati, via Enrico Fermi 40, 
00044 Rome, Italy \\
e-mail: Gaia.Lanfranchi@lnf.infn.it}

\maketitle

\abstract{
Recent results obtained by the KLOE experiment operating at DA$\Phi$NE, the
Frascati $\phi$-factory, are presented.
They mainly concern neutral kaon decays including the $K_L$ dominant branching ratios, the $K_L$ lifetime and the extraction of the $CKM$ parameter 
$V_{us}$ from the $K_L$ semileptonic decays and lifetime.
The best world upper limit on $K_S \to \pppo$ channel is also presented.}

\section{Introduction}
\label{sec:intro}

The determination of $|V_{us}|$ and $|V_{ud}|$ provide the most
precise test of CKM unitarity. 
In fact the first row unitarity requires $|V_{ud}|^2+|V_{us}|^2+
|V_{ub}|^2 = 1$ which, since $|V_{ub}|^2 \sim 10^{-5}$, is equivalent
to $|V_{ud}|^2+|V_{us}|^2 = 1$.
The 2004 edition of Particle Data Group \cite{pdg2004} gives
$|V_{ud}| = 0.9738 \pm 0.0005$ and $|V_{us}| = 0.2200 \pm 0.0026$
from which the sum of the squares gives $0.9966 \pm 0.0015$
which deviates from unitarity by $\sim 2 \sigma$.
Semileptonic kaon decays are the cleanest way to obtain an accurate value 
of $|V_{us}|$. In fact, since  $K \to \pi$ is a $0^- \to 0^-$ transition,
only the vector part of the weak current has a non vanishing contribution
and such processes are protected by the Ademollo-Gatto 
theorem against SU(3) breaking corrections to the lowest order in $m_s - m_d$.

 In order to extract $|V_{us}|$ from semileptonic kaon decays we need to 
 measure the branching fraction and the $K_L$ lifetime.
 The most accurate determination of $|V_{us}|$ from $K_L$ semileptonic decays
 comes from KTeV collaboration \cite{ktev:br}.  
 The $K_{e3}$ branching fraction is found to be $0.4067 \pm 0.0011$, strikingly different from PDG fit value, $0.3881 \pm 0.0027$.
 The KTeV result has been partially confirmed by the 
 NA48 Collaboration\cite{na48:br}, $0.4010 \pm 0.0045$, 
 even if with an error $\sim$ 4 times bigger.
 
 The $K_L$ lifetime value in the PDG relies on a single direct measurement
  $\tau_L = (51.54 \pm 0.44)$ ns 
  that was performed more than 30 years ago
 \cite{vosburgh:tkl}. $K_L$ lifetime is, at present, the major experimental 
  source of error in the determination of $|V_{us}|$.

 KLOE has the unique possibility to measure simultaneously $K_L$ absolute
 branching fractions and lifetime. 
 The measurement of the $K_L$ absolute branching is possible
 at the $\phi$-factory since the production of $K_S K_L$ pairs
 in $\phi$ decays provides a tagged, monochromatic $K_L$ beam of known flux.
 Moreover $K_L$ have low momentum ($|\vec{p}_{K_L}| \sim $ 110 MeV/c)
and, therefore, a big fraction ($\sim 50\%$) of them decays inside the detector. 

 This is important since the statistical error on the lifetime 
 depends on the number of events, $\propto 1/\sqrt{N}$ and, very strongly, 
 on the time interval covered:

\begin{equation} 
{\delta \tau \over \tau }(\mbox{stat}) = {\delta \Gamma \over \Gamma } = {1 \over \sqrt{N}} \times 
\left [ {-1 + e^{3T} + (e^T-e^{2T})(3+T^2) \over (-1+e^T)^3 }\right]^{-0.5}
\label{eq:paolo} 
\end{equation}

\noindent
where $T=\delta t/\tau$ is the interval covered by the fit
in lifetime units and $N$ is the number of events in that interval.
With $T \sim 0.4 $ and $N \sim 9 \cdot 10^6$ (which is the KLOE case)
a statistical error of $\sim$ 0.3$\%$ can, in principle, be reached.

\section{Experimental Setup}
\label{sec:setup}

DA$\Phi$NE, the Frascati $\phi$ factory, is an $e^{+}e^{-}$ collider
working at $W \sim m_{\phi} \sim 1.02$ GeV with a design luminosity
of $5 \times 10^{32}$ cm$^{-2}$ s$^{-1}$. The $\phi$ mesons are produced,
almost at rest, with a visible cross section of $\sim$ 3.2 $\mu$b
and decay into $K^+ K^-$ ($K_S K_L$) pairs with BR of $\sim 49$\% 
($\sim 34$\%).
These pairs are produced in a pure $J^{PC}=1^{--}$ quantum state, so that 
observation of a $K_S$ ($K^+$)  in an event signals (tags) 
the presence of a $K_L$ ($K^-$) and vice-versa; 
highly pure and nearly monochromatic $K_S,K_L,K^+$ and $K^-$
beams can thus be obtained. Neutral kaons get a momentum of $\sim$ 110 MeV/c which
translates in a slow speed, $\beta_{K} \sim$ 0.22.
$K_S$ and $K_L$ can therefore be distinguished by their mean decay lengths:
$\lambda_{S} \sim $ 0.6 cm and $\lambda_{L} \sim $ 340 cm.
 
The KLOE detector consists of a drift chamber, DCH, surrounded by an
electromagnetic calorimeter, EMC. 
The DCH~\cite{nimdch} is a cylinder of 4 m diameter
and 3.3 m in length which constitutes a large fiducial volume 
for $K_L$  decays (1/2 $\lambda_{L}$). The momentum resolution for tracks 
at large polar angle is $\sigma_{p}/p \leq 0.4$\%. The EMC is a 
lead-scintillating fiber calorimeter~\cite{nimcalo}
consisting of a barrel and two endcaps which cover 98\% of the solid angle. The
energy resolution is $\sigma_{E}/E \sim 5.7\%/\sqrt{\rm{E(GeV)}}$. 
The intrinsic time resolution is $\sigma_{T} =$ 54 ps$/\sqrt{\rm{E(GeV)}} \oplus 50$ ps. 
A super-conducting coil surrounding the barrel 
provides a 0.52 T magnetic field.

During 2002 data taking, the maximum luminosity reached by DA$\Phi$NE was 
7.5$\times 10^{31}$ cm$^{-2}$ s$^{-1}$. Although this is lower than the design value,
the performance of the machine was improving during the years and, at the end
of 2002, we collected $\sim$ 4.5 pb$^{-1}$/day. The whole data sample in the years 2001-2002
amounts to 450 pb$^{-1}$, equivalent to 1.4 billion $\phi$ decays. 
The analyses presented here are based on $\sim$ 400 pb$^{-1}$ of integrated luminosity
of the 2001 and 2002 runs.

Recently, 
the machine has been upgraded and KLOE has resumed its data taking in 
April 2004. Up to know (15$^{th}$ May, 2005) $\sim$ 1 fb$^{-1}$ have 
already been collected with a peak luminosity  of 
1.3$\times 10^{32}$ cm$^{-2}$ s$^{-1}$. 
We foresee to reach $\sim$ 2 fb$^{-1}$ by the end of the year.

\section{Measurement of the dominant $K_L$ branching ratios}
\label{sec:abr}


The $K_L$ absolute branching fractions are determined 
on a tagged $K_L$ events sample by counting the number
of $K_L$ decays in each channel, $N_i$, in the used fiducial volume and 
correcting for acceptance, reconstruction efficiency 
and tagging efficiency:

\begin{equation}
  BR(K_L \to i) = { N_i \over N_{tag} } \times 
 {1 \over \epsilon_{rec}(i)\times \epsilon_{FV}(\tau_L) \times
 \epsilon_{tag}(i)/\epsilon_{tag}(all) }
\label{eq:br}
\end{equation}

\noindent where $i=\pppo, \pen, \pmn, \pppco$, $N_{tag}$ is the number of tagging events, $\epsilon_{rec}$ is the reconstruction efficiency 
($\sim 45\%$ for $\pppco$ events, $\sim 60\%$ for $\pen, \pmn$ events 
and $\sim 100\%$ for $\pppo$ events), $\epsilon_{FV}(\tau_L) $ 
is the fiducial volume geometrical acceptance
which depends on the $K_L$ lifetime and 
$\epsilon_{tag}(i)/\epsilon_{tag}(all)$
is the fractional variation of the tagging efficiency when
the $K_L$ decays in a given channel with respect to the average. 
We define this ratio {\em tag bias}.

\subsection{The tag}
\label{subsec:tag}

The tag is provided by $K_S \to \ppc$ events selected 
requiring the presence of a vertex with two opposite curvature tracks 
within a cylinder of radius $r<$ 10 cm and height $h<$ 20 cm 
around the interaction point (IP). 
The two-tracks invariant mass, in the pions hypothesis, must be within 5 MeV around 
$m_{K_S}$. The magnitude of the total momentum of the two tracks 
must be within 10 MeV of the value expected from the value of $\vec{p}_{\phi}$.
The $K_L$ momentum is obtained from the decay kinematics of $\phi \to K_S K_L$
using the $K_S$ direction reconstructed from the measured momenta 
of $\pi^+ \pi^-$ tracks and the known value of $\vec{p}_{\phi}$.

The main source of tag bias is due to the dependence of the trigger efficiency 
on the $K_L$ behaviour. The hardware calorimeter 
trigger which requires two local energy deposits above 
some thresholds (50 MeV on the barrel and 150 MeV on the end caps) 
is used for the present analysis.
The trigger efficiency is essentially 100$\%$ for $\pppo$, between $95-85\%$ for charged 
decays.
To reduce the tag bias due to trigger efficiency 
we require that the trigger conditions are satisfied by the pions from
$K_S$. We further reinforce these conditions by requiring that the two pions 
impinge on the calorimeter barrel and produce two clusters 
with an energy $E \geq $ 80 MeV each.
 
The $FV$ used for the analysis is defined inside the drift chamber 
by 35 cm $< \sqrt{x^2+y^2} <$ 150 cm and $|z| <$ 120 cm, where $(x,y,z)$ are the $K_L$ decay vertex position coordinates. Since the $K_L$ mean decay path length in KLOE is $\sim$ 340 cm, the FV contains $\sim 26.1\%$ of the $K_L$ decays.
This choice minimize the difference in tag bias among the decay modes. The average tag bias is 0.985, 0.99 and 1.02 for $\pen$ or $\pmn, \pppco$ and
$\pppo$ decays respectively. 

\subsection{Charged $K_L$ decays}
\label{subsec:klcharged}

The $K_L$ in charged decay modes are selected by requiring the presence 
of two good tracks forming a vertex in the FV 
and not belonging to the $K_S$ decay tree.
A track is associated to the $K_L$ if the 
point of closest approach to the $K_L$ line of flight 
has a distance with respect to the $K_L$ line of flight 
$d_c < a \sqrt{x^2+y^2}+b$, with $a=0.03$ and $b$= 5 cm.

The tracking efficiency has been determined by counting the number of events
with at least one found $K_L$ track and the number of events
in which there are two opposite sign decay tracks.
The tracking efficiency is also evaluated from Monte Carlo 
simulation and it is 60.5$\%$ for $K_{e3}$, 58.5$\%$ 
for $K_{\mu 3}$ and $43.0 \%$ for $\pppco$. 
A correction to the tracking efficiency 
has been applied by comparing the results for data and Monte Carlo simulation.
The correction is evaluated as a function of the track momentum using 
$K_L \to \pppco$ and $K_L \to \pen$ events and ranges 
between 1.03 and 0.99 depending on the channel.

\begin{figure}[hbt]
\begin{center}
\begin{tabular}{c}
\centering \epsfig{figure=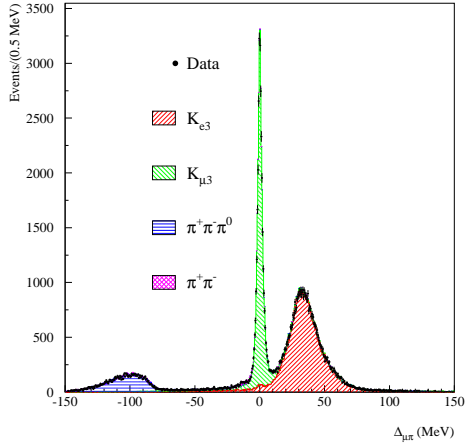,width=7cm}
\end{tabular}

\caption{$\Delta_{\pi \mu}$ distribution for an event subsample.
Dots are data, solid histograms are Monte Carlo. 
\label{fig:shapes}}
\end{center}
\end{figure}

The variable to discriminate among the different charged decay modes
is the lesser of the two values of 
$\Delta_{\pi \mu} = |\vec{p}_{mis}| - E_{miss}$, where $\vec{p}_{mis}$ is the 
missing momentum  and $E_{miss}$ is the missing energy 
evaluated in the two mass assignments $\pi^+, \mu^-$ or $\pi^- \mu^+$.
An example of this distribution is shown in Fig.~\ref{fig:shapes} where
the different components are shown.

\begin{figure}[t]
\vspace{0.4cm}
\begin{center}
\begin{tabular}{cc}
\begin{minipage}{0.5\linewidth}
    \centering \epsfig{file=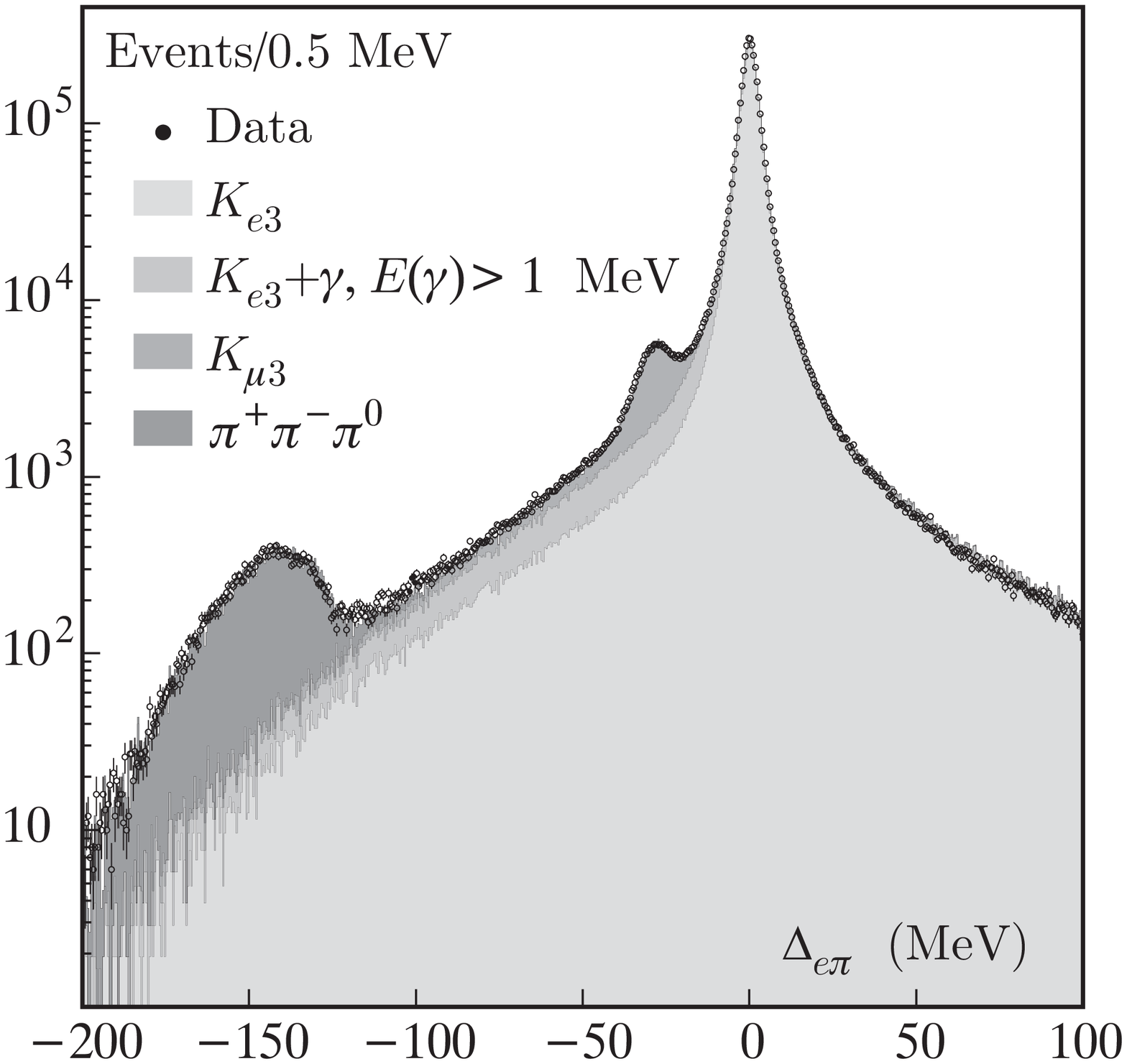,width=6cm} 
\end{minipage}
&
\begin{minipage}{0.5\linewidth}
    \centering \epsfig{file=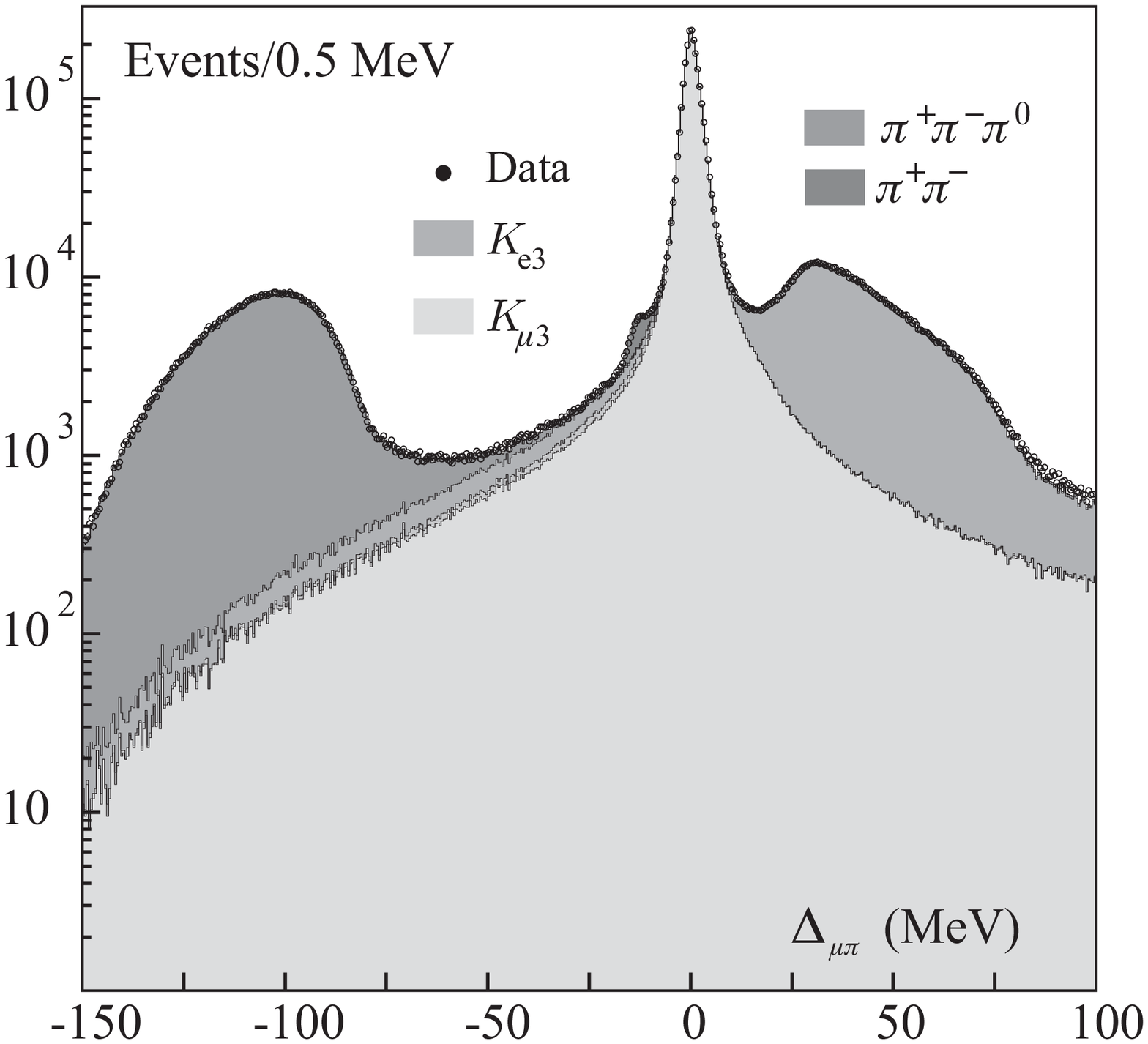,width=6cm}  
\end{minipage}
\\
\end{tabular}
\caption{$\Delta_{\pi e}$ (left) and $\Delta_{\pi \mu}$ (right)
  distributions for data (dots) and Monte Carlo (solid histograms).
\label{fig:shapes_log}}

\end{center}
\end{figure}

The $\Delta_{\pi \mu}$ distribution obtained with data 
is fitted with a linear combination of three Monte Carlo
distributions (for $\pen, \pmn$ and $\pppco$ events) by leaving free the
relative weights. 
The contribution from the $CP$ violating decay $K_L \to \pi^+ \pi^-$ 
and the $K_L \to \pppo$ with Dalitz conversion is kept
fixed in the fit and amounts to 0.3 $\%$.
In Fig.~\ref{fig:shapes_log} we show enriched samples of 
$K_{e3}$ (left) and $K_{\mu 3}$ (right) events obtained by selecting the 
$e$ or $\mu$ by time of arrival and energy deposition in the calorimeter.
The radiative corrections which affect 
mainly the $K_{e3}$ decays have been properly taken 
into account in the Monte Carlo
simulation~\cite{kn194}.

\subsection{Neutral $K_L$ decays}
\label{subsec:klneutral}

\begin{figure}
\vspace{0.4cm}
\begin{center}
\begin{tabular}{c}
\begin{minipage}{0.5\linewidth}
    \centering \epsfig{file=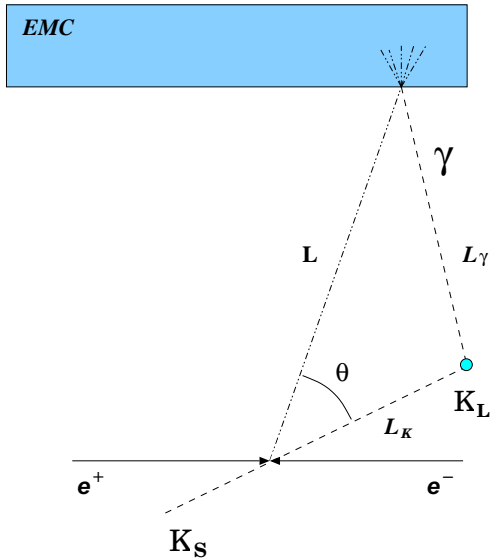,width=7cm} 
\end{minipage}
\end{tabular}
\caption{ Left: the time of flight triangle. \label{fig:lk}}

\end{center}
\end{figure}

$K_L \to \pppo$ decays are selected using time of flight techniques.
In fact, the position of the $K_L$ vertex for 
$K_{L} \to \pi^{0} \pi^{0} \pi^{0}$ decays is measured using the photon
arrival times on the EMC.
Each photon defines a time of flight
triangle shown in Fig.\ref{fig:lk}. The three sides are the $K_{L}$
decay length, $L_{K}$; the distance from the decay vertex to the
calorimeter cluster centroid, $L_{\gamma}$; and the distance from
the cluster to the $\phi$ vertex, $L$. The equations to determine
the unknowns $L_K$ and $L_{\gamma}$ are:

 \begin{eqnarray}
   L^2 + L_K^2 -2 L L_K \cos\theta & = & L_{\gamma}^2 \\ \nonumber
   L_K/\beta_K + L_{\gamma} & = & ct_{\gamma}
 \label{eq:tr} 
 \end{eqnarray} 

\noindent
where $t_{\gamma}$ is the photon arrival time on the EMC, 
$\beta_{K} c $ is the $K_{L}$ velocity and $\theta$ is 
the angle between $\vec{L}$ and $\vec{L}_{K}$. 
Only one of the two solutions is kinematically correct.
The $K_L$ vertex position is obtained by the energy weighted average
of each $L_K$ measurement.

The accuracy of this method is checked with $K_L \to \pppco$ 
decays, comparing the position of the $K_L$ decay 
vertex from tracking using the $\ppc$ pair with the one from 
timing with the two photons from $\pi^0$.
The vertex reconstructed by the calorimeter has on average an offset
of 2 mm almost uniform in the fiducial volume.

To select the $K_L \to \pppo$ events we require at least three
photons with energy greater than 20 MeV originating from the same vertex.

The main sources of inefficiencies are: 1) geometrical acceptance;
2) cluster energy threshold; 3) merging of clusters; 4) accidental
association to a charged track; 5) Dalitz decay of one or more
$\pi^{0}$'s. The effect of these inefficiencies is to modify the
relative population for events with 3, 4, 5, 6, 7 and $\ge 8$,
clusters with a loss of efficiency of $\sim 0.8\%$.

Background contamination affects only events with three and four clusters.
The main source of background 
comes from $K_L \to \pppco$ decays where one or two
charged pions produce a cluster not associated to a track 
and neither track is associated to the $K_L$ vertex.
This background is rejected by requiring at least one cluster
with at least 50 MeV energy and a polar angle satisfying 
$|\cos \theta | < 0.88$.
Other sources of background are  $K_{L} \to
\pi^{0} \pi^{0}$ decays, possibly in coincidence with machine
background particles ($e^{\pm}$ or $\gamma$) that shower in the
QCAL and generate soft neutral particles and $ K_S \to \pi^0 \pi^0$ 
following the $K_L \to K_S$ regeneration in the drift chamber material.

\subsection{Results}
\label{subsec:klbr}

A total of $\sim 13$ milions of tagged $K_L$ are used to compute
the branching fractions, almost $\sim$ 40 milions to evaluate
systematic uncertainties. Using the published result of the $K_L$ 
lifetime ($\tau_L = 51.54 \pm 0.44$ ns)\cite{vosburgh:tkl} we obtain the following results:

\begin{eqnarray}
 BR(K_L \to \pen)   &=& 0.4049 \pm 0.0010_{stat} \pm 0.0031_{syst} \\ \nonumber
 BR(K_L \to \pmn)   &=& 0.2726 \pm 0.0008_{stat} \pm 0.0022_{syst} \\ \nonumber
 BR(K_L \to \pppo)  &=& 0.2018 \pm 0.0004_{stat} \pm 0.0026_{syst} \\ \nonumber
 BR(K_L \to \pppco) &=& 0.1276 \pm 0.0006_{stat} \pm 0.0016_{syst} \\ 
\end{eqnarray}

\noindent where the sources of systematic uncertainties are shown
in Table \ref{tab:syst}. The sum of all measured branching fraction above,
plus the PDG value for rare decays, 0.0036, 
is $\sum BR_i = 1.0104 \pm 0.0018_{\mbox{correlated}} \pm 0.0074_{\mbox{uncorrelated}}$
where the correlated error includes all contributions 
to the uncertainties on the branching ratios that are 100$\%$ correlated
between channels, such as the uncertainty in the value of the $K_L$
lifetime.
This result depends on the value of the $K_L$ lifetime through the
acceptance (Eq. \ref{eq:br}). 
Turning the argument around, by normalizing the sum to 1 we obtain un
indirect estimate of the $K_L$ lifetime:

\[
  \tau(K_L) = (50.72 \pm 0.14_{\mbox{stat}} \pm 0.36_{\mbox{syst}}) \mbox{ns}
\]

\noindent and a new set of values for the branching fractions:

\begin{eqnarray}
 BR(K_L \to \pen)   &=& 0.4007 \pm 0.0006_{stat} \pm 0.0014_{syst} \\ \nonumber
 BR(K_L \to \pmn)   &=& 0.2698 \pm 0.0006_{stat} \pm 0.0014_{syst} \\ \nonumber
 BR(K_L \to \pppo)  &=& 0.1997 \pm 0.0005_{stat} \pm 0.0019_{syst} \\ \nonumber
 BR(K_L \to \pppco) &=& 0.1263 \pm 0.0005_{stat} \pm 0.0011_{syst} \\ 
\end{eqnarray}

\begin{table}[t]
\vspace{0.4cm}
\begin{center}
\begin{tabular}{|l|c|c|c|c|}
\hline
            & $\pen$  &  $\pmn$  &  $\pppco$   &  $\pppo$  \\ \hline
Selection   & 0.0011  &  0.0007  &  0.0004     &  0.0020   \\
Signal Shape       & 0.0006  &  0.0009  &  0.0010     &   -       \\ 
Tag Bias    & 0.0013  &  0.0008  &  0.0007     & 0.0005    \\
lifetime    & 0.0023  &  0.0017  &  0.0007     & 0.0012    \\ \hline
\end{tabular}
\caption{ Summary of systematic uncertainties on the absolute
branching fractions measurements. \label{tab:syst}}
\end{center}
\end{table}

\section{Direct measurement of the $K_L$ lifetime}
\label{sec:tau}

 We have measured the $\kl$ lifetime using $\sim 15 \times 10^6$ events of
 the fully neutral decay $\kl \to \pppo$ tagged by 
 $\ks \to \pi^+ \pi^-$ events. This choice is motivated by the fact that
 we want to maximize the number of tagged events to reduce the
 statistical error and, simultaneously, we want to 
 avoid any coupling among tagging and tagged events 
 in order to minimize the systematic uncertainty.

 The $\kl \to \pppo$ sample selected for the BR measurement has been used
 also for the $K_L$ lifetime measurement plus some additional cuts.


 For lifetime measurement we must keep under control 
the variation of the selection efficiency with $L_K$. 
 Monte Carlo simulation shows that the selection efficiency 
 has a linear dependence with $L_K$, 
 $\epsilon(L_K) = (0.9921 \pm 0.002) - (1.9 \pm 0.2) \cdot 10^{-5}\cdot L_K$(cm)
 mainly due to the vertex reconstruction efficiency.
 The vertex reconstruction efficiency as a function of $L_K$ has been checked 
 also using $\kl \to \pppco$ events both in data and in Monte Carlo simulation.
 We find the same linear dependence as in the $\kl \to \pppo$ case with slopes
 compatibles within their statistical uncertainties.

 The $K_L$ lifetime is measured using events with
 a vertex reconstructed in the region
 40 cm $< L_K <$ 165 cm and with a flight direction defined by a 
 polar angle $\theta$
 with respect to the beam axis between $40^{\circ} < \theta < 140^{\circ}$.
 These two conditions define the fiducial volume.

 The $\kl$ proper time, $t^*$, is obtained event by event dividing 
 the decay length $L_K$ by $\beta \gamma$ of the $\kl$  in the laboratory,
 $t^* = L_K/(\beta\gamma c)$.
 The residual background is subtracted bin by bin using Monte Carlo predictions.

 The variations of the vertex reconstruction efficiency 
 as a function of the decay length are taken into account by correcting 
 bin by bin the decay vertex distribution with the efficiency 
 values obtained with 
 the $\kl \to \pppo$ Monte Carlo sample multiplied by the data - Monte Carlo ratio
 of the efficiencies evaluated with $\pppco$ sample.

 The statistical uncertainty of the efficiency values 
($\sim 0.1\%$)
 has been taken into account by adding it in quadrature to the statistical fluctuation
 of the entries in each bin of the $t^*$ distribution after the background subtraction. 

 The distribution is fitted with an
 exponential function
 inside the fiducial volume, 
 which, in terms of proper time ranges from 6 ns to 24.8 ns. This corresponds
 to a time interval $T \sim 0.37$ expressed in lifetime units.  
 With $\sim 8.5 \cdot 10^6$ events inside the fit region we obtain
$ \tau = (50.87 \pm 0.17_{\mbox{stat}})$ {ns}
with a $\chi^2$ = 58 for 62 degrees of freedom. 
The major sources of systematic uncertainties come from the background
evaluation, tagging and selection efficiency and from the estimate of
the $K_L$ nuclear interactions in the drift chamber material.
The total systematic error is $\sim 0.5\%$. Our result is:
\[
  \tau(K_L) = (50.87 \pm 0.17_{\mbox{stat}} \pm 0.25_{\mbox{syst}}) \mbox{ns}
\]
It is compatible at 1.3 $\sigma$ level with the other
measurement\cite{vosburgh:tkl} and only at 1.7$\sigma$ level with the PDG
2004 fit \cite{pdg2004}.

\begin{table}[t]
\vspace{0.4cm}
\begin{center}
\begin{tabular}{cc}
\begin{minipage}{0.5\linewidth}
    \centering \epsfig{file=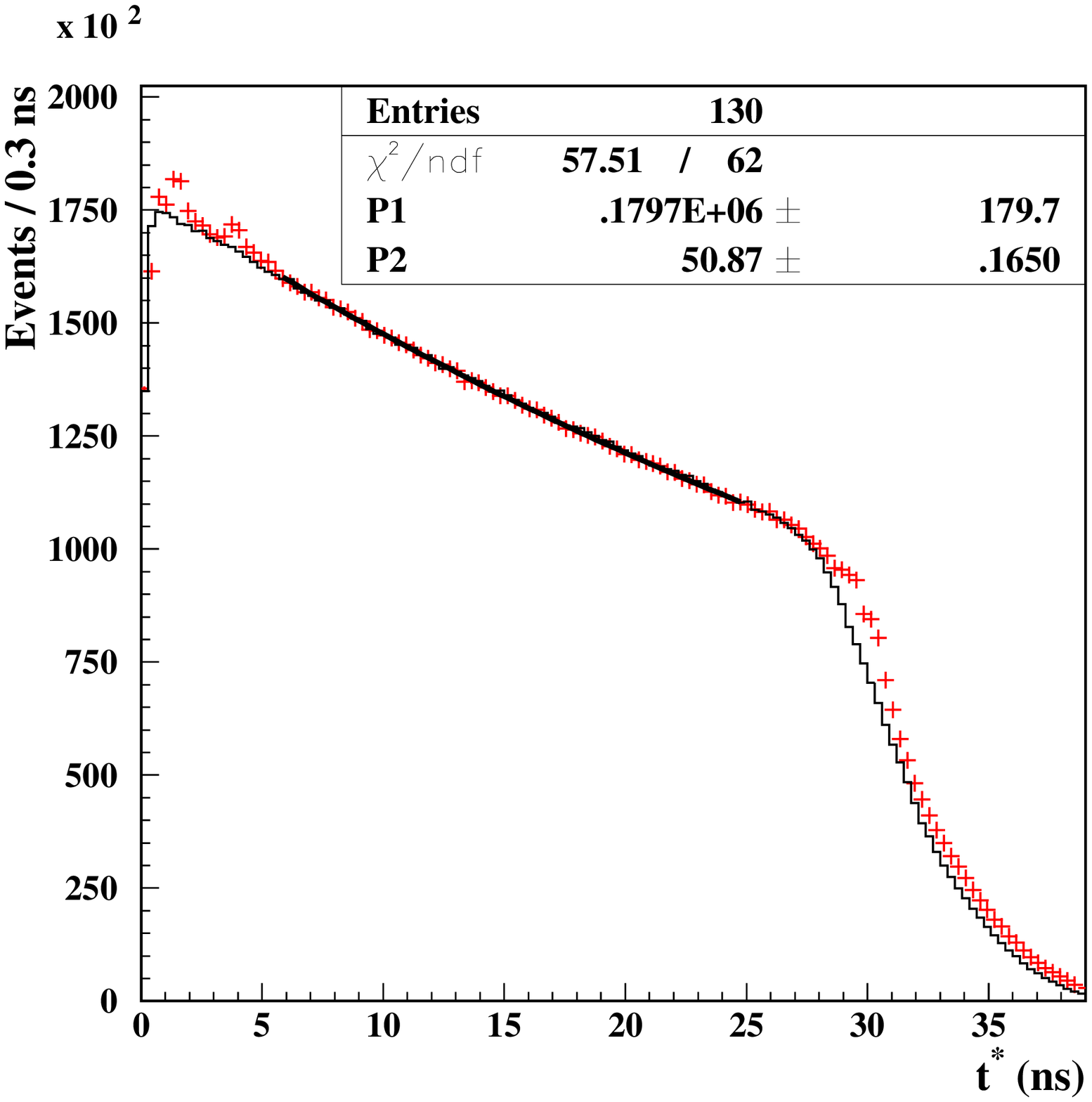,width=6cm}
\end{minipage}
&
\begin{minipage}{0.5\linewidth}
    \centering \epsfig{file=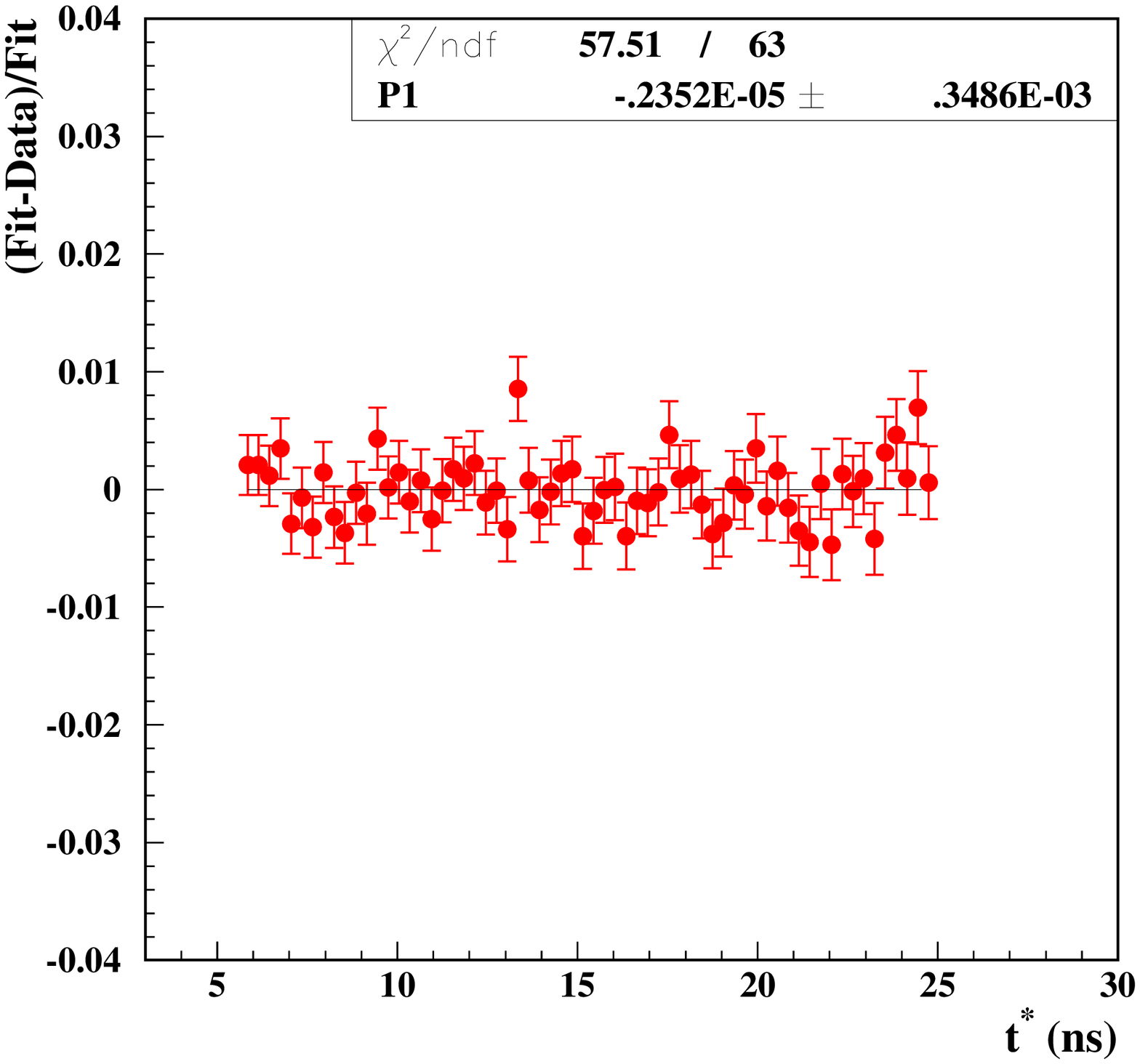,width=6cm}
\end{minipage}
\\
\end{tabular}
\caption{ Left: 
fit of the proper time distribution. Right: residuals of the fit.
\label{tab:exp}}
\end{center}
\end{table}

\section{Determination of $|V_{us}|$}
\label{sec:tau}

\begin{figure}[htb]
    \centering \epsfig{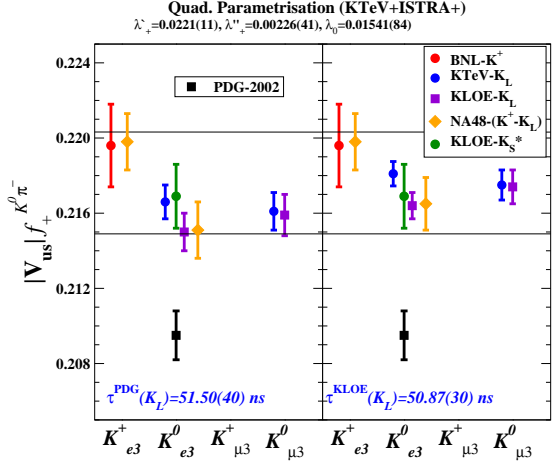}
\caption{\small $|V_{us}|\times f_{+}^{K^0 \pi}(0)$ for various $K$
  semileptonic decays.}
 \label{fig:vus}
\end{figure}

$|V_{us}|$ is proportional to the square root 
of the semileptonic BR of $K$ mesons.
For $K_L^{e3}$ decays we can write~\cite{CKMwsp}:

\begin{equation}
|V_{us}|\times f_{+}^{K^0 \pi}(0) 
= \left[ { 128 \pi^3 BR(K_{L} \to \pi e \nu) \over
 \tau_{L} G_{\mu}^2 M^5 S_{ew} I_K^e(\lambda_+, \lambda'_+) }
\right ]^{1/2} \times { 1 \over 1+\delta^{Ke}_{\rm em} } 
\label{eq:vus}
\end{equation}

\noindent where $f_{+}^{K^0}(0)$ is the vector form factor
 at zero momentum transfer and 
$I_i(\lambda_+,\lambda'_+) $ is the integral of the phase space density, 
factoring out $f_+^{K^0 \pi}$ and without radiative corrections.
Radiative corrections at large scale of form factor and phase space density 
are contained in the term 
$\delta_{em}^{Ke} = (0.55 \pm 0.10) \%$ ~\cite{cir1,cir2}. 
The short-distance electroweak corrections are included in the parameter 
$S_{ew}$.
$\lambda_+$ and $\lambda'_+$ are the slope and 
curvature of the vector form factor $f_+^{K^0 \pi}(0)$. 
From eq. \ref{eq:vus} we see that the lifetime value enters directly
in the determination of the product $|V_{us}|\times f_{+}^{K^0 \pi}(0)$.
The three most recent results are:

 \[
  \begin{array}{ll}
         BR(\kl \to \pen) & = (40.67 \pm 0.11) \% \;\;\; \mbox{KTeV} \\
         BR(\kl \to \pen) & = (40.10 \pm 0.45) \% \;\;\; \mbox{NA48} \\
         BR(\kl \to \pen) & = (40.07 \pm 0.15) \%  \;\;\; \mbox{KLOE} 
   \end{array}
\]  
where the error is the sum in quadrature of the statistical and 
systematic uncertainties.
Fig. \ref{fig:vus} (left) shows the product $|V_{us}|\times f_{+}^{K^0 \pi}(0)$
extracted from the three recent measurements of $BR(\kl \to \pen)$ using 
the PDG 2004 average $\tau = (51.50 \pm 0.40)$ ns.

In the extraction of $V_{us}$
 we use the values of $\lambda_+$ and $\lambda'_+$ 
obtained by KTeV experiment from a quadratic fit~\cite{ktev:ff}.
In the same plot we show also the value 
of $|V_{us}|\times f_{+}^{K^0 \pi}(0)$ obtained from 
the KLOE preliminary measurement~\cite{kloe:lathuile} of $BR(\ks \to \pen) = (7.09 \pm 0.07_{stat} \pm 0.08_{syst})\cdot 10^{-4}$ 
and the $\pm 1 \sigma$ band from the $|V_{ud}|$ and unitarity where we use
$f_+^{K^0 \pi} (0) = 0.961 \pm 0.008$ following Ref. ~\cite{f0}.
Fig.~\ref{fig:vus} (right) shows the same product extracted using the
KLOE $\kl$ lifetime value, $\tau_L = (50.87 \pm 0.17 \pm 0.25)$ ns.
The $\kl$ data are now in better agreement with $\ks$ ones and unitarity.

\section{Direct search of $K_S \to \pppo$ decay}
\label{sec:ks3p0}

The decay $K_S \to 3 \pi^0$\  is a pure CP violating process. 
The related CP violation parameter $\eta_{000}$ is defined as the 
ratio of $K_S$\ to $K_L$\ decay amplitudes:
$\eta_{000} = A(K_S \to 3 \pi^0)/ A( K_L \to 3 \pi^0)=
\varepsilon + \varepsilon^{'}_{000}$
where $\varepsilon$ describes the CP violation in the mixing matrix 
and $\varepsilon^{'}_{000}$ is a direct CP violating term. 
In the standard model we expect  $\eta_{000}$ to 
be similar to  $\eta_{00}$ ($|\eta_{00}| \sim 2 \times 10^{-3}$). 
The expected branching ratio of this decay
is therefore $\sim 2 \times 10^{-9}$, making its direct 
observation really challenging.
The best upper limit on the BR (i.e. on $|\eta_{000}|^2$) has been set 
to 1.4$\times 10^{-5}$ by SND\cite{SND3pi0} where, 
similarly to KLOE,  it is possible to tag a $K_S$ beam. 

The other existing technique
is to detect the interference term between $K_S K_L$ in the same
final state which is proportional to $\eta_{000}$,
$\Re \eta_{000} \cos(\Delta m t) - \Im \eta_{000}  \sin(\Delta m t)$.
The best published result
using this method comes from the NA48 Collaboration\cite{na48:ks000}.
Fitting the $K_S-K_L \to \pppo$ interference pattern at small decay times, 
they find 
$\Re \eta_{000} = - 0.002 \pm 0.011_{\mbox stat} \pm 0.015_{syst}$ and
$\Im \eta_{000} = - 0.003 \pm 0.013_{\mbox stat} \pm 0.017_{syst}$
corresponding to $BR(K_S \to 3\pppo) \leq 7.4 \times 10^{-7}$ at 
$90 \%$ C.L.

The signal selection requires a $K_L$ interacting with the calorimeter 
and six neutral clusters coming 
from the interaction point (IP). A first rejection of the huge background 
coming from the decay $K_S \to \pi^0 \pi^0$ + 2 fake $\gamma$ 
is obtained by applying  a kinematic fit imposing as constraints
the $K_S$ mass, the $K_L$ 4-momentum and $\beta=1$ for each photon. 
 Two pseudo-$\chi^2$ variables are then built: 
$\chi^2_{3\pi^0}$ which is based on
the best $6\gamma$ combination into $\pppo$
and $\chi^2_{2\pi^0}$ which selects four out of six photons providing the best 
kinematic agreement with the $K_S \to \pi^0 \pi^0$ decay.
A signal box is defined in the $\chi^2_{2\pi^0}$ vs $\chi^2_{3\pi^0}$ 
plane by optimising the upper limit in the Monte Carlo sample.
Residual background comes from $K_S \to \pi^+\pi^-, K_L \to 3 \pi^0$ events in which one 
of the two pions interacting with the quadrupoles produces a 
late cluster  simulating a $K_L$-crash.
This background is rejected by vetoing events with two charged tracks from IP. 

At the analysis end we find 2 events in the signal box with an estimated
background of $B = 3.13 \pm 0.82_{\rm stat} \pm 0.37_{syst}$.
To derive the upper limit on the number of signal counts, we build the
background probability distribution function, taking into account our
finite MC statistics and the uncertainties on the MC calibration factors.
This function is folded with a Gaussian of width equivalent to the entire
systematic uncertainty on the background. Using the Neyman construction
\cite{neyman} we limit the number of $K_S \to \pppo$ decays observed to
3.45 at 90$\%$ C. L. with a total reconstruction efficiency of
$(24.36 \pm 0.11_{\rm stat} \pm 0.57_{\rm syst})\%$. In the same tagged
sample, we count $3.78 \times 10^7$ $K_S \to \ppo$ events.
We use them as normalization. Finally, using the value 
$BR(K_S \to \ppo) = 0.3105 \pm 0.0014$ \cite{pdg2004} we obtain
$ BR(K_S \to \pppo) \leq 1.2 \times 10^{-7}$ at 90 \% C.L. which represents
an improvement by a factor $\sim$ 6 with respect to the best previous 
limit\cite{na48_ks3p0}.



\end{document}